\documentclass[11pt]{article}

\usepackage[margin=1in]{geometry}
\usepackage{graphicx}
\usepackage{amsmath}
\usepackage{booktabs}
\usepackage{siunitx}
\usepackage[numbers,sort&compress]{natbib}
\usepackage{hyperref}
\usepackage{authblk}

\title{From DAMPE to LHAASO: Rigidity Scales and Composition Changes in Galactic Cosmic Rays}

\author[1]{Ruo-Yu Liu}
\author[2]{Zhen Cao}

\affil[1]{School of Astronomy and Space Science, Nanjing University, 210023 Nanjing, China}

\affil[2]{State Key Laboratory of Particle Astrophysics \& Experimental Physics Division \& Computing Center,\\
Institute of High Energy Physics, Chinese Academy of Sciences, 100049 Beijing, China}

\begin{document}

\maketitle

Cosmic rays (CRs) are high-energy particles that reach Earth after travelling through the Galaxy. Most of them are nuclei, from protons to heavier elements such as carbon, oxygen, and iron. Their spectra carry information about where they are accelerated and how they propagate through Galactic magnetic fields. For a long time, the spectrum below the knee was often treated as close to a single power-law. Recent measurements have shown that this picture is too simple. The spectra of individual species show structures, including a hardening at a few hundred GV, where the spectrum becomes flatter and the flux decreases more slowly with rigidity,  and a softening at higher rigidities, where the spectrum becomes steeper and the flux decreases more rapidly.

The new Dark Matter Particle Explorer (DAMPE) measurement of carbon, oxygen, and iron spectra gives a clearer view of these structures \citep{DAMPE2026}. Using nine years of data, DAMPE measured these primary nuclei, which are mainly accelerated at astrophysical sources, up to about 100\,TV in rigidity, and up to about 60\,TV for iron. This is not an easy task. The event rate of these nuclei is low, the energy reconstruction requires unfolding, and the detector response depends on hadronic interaction models. These difficulties are especially important for heavy nuclei at the highest energies, where the flux is low and the statistics become limited. Even so, together with updated proton and helium spectra, the DAMPE data suggest that all five species soften at nearly the same rigidity, around 15\,TV. This points to a possible common rigidity scale.

The result is important because rigidity is the quantity that controls the motion of charged particles in magnetic fields. It is defined as $R=pc/Ze$, where $p$ is momentum, $c$ is the speed of light, $Z$ is the particle charge number, and $e$ is the elementary charge. If a spectral break is caused by an acceleration limit or by a change in propagation, different nuclei may show the break at the same rigidity. In the energy spectrum, this means that the break energy should scale with charge $Z$. If the break were instead caused by a mass-dependent interaction effect, one might expect a scaling closer to mass number $A$.

The origin of this common softening in the rigidity spectra of primary CRs remains open. It could reflect an additional source component, such as CRs from a nearby supernova remnant superposed on a smoother Galactic background. It could also be a propagation effect. If the diffusion coefficient changes its rigidity dependence near 15\,TV, primary nuclei may acquire similar spectral breaks during transport through the interstellar medium \citep{Strong2007}. These two pictures may look similar in primary spectra alone, but they can differ in other observables.  Secondary nuclei, such as boron, lithium, and beryllium, are mainly produced when heavier primary nuclei collide with interstellar gas. A secondary-to-primary
ratio, such as B/C or B/O, compares the flux of such a secondary species with that of a primary species and is therefore sensitive to propagation. If the 15\,TV feature is mainly caused by propagation, related structures should appear at a similar rigidity in these ratios. If
it is mainly caused by a local source that adds primary particles, the response in the secondary spectra should be weaker, shifted, or absent, depending on the source age, distance, composition, and the amount of secondary production near the source. Current data do not yet allow a firm distinction between these two scenarios. In this sense, composition and secondary nuclei are not auxiliary checks, but part of the main test of the DAMPE interpretation.

\begin{figure}[t]
    \centering
    \includegraphics[width=\linewidth]{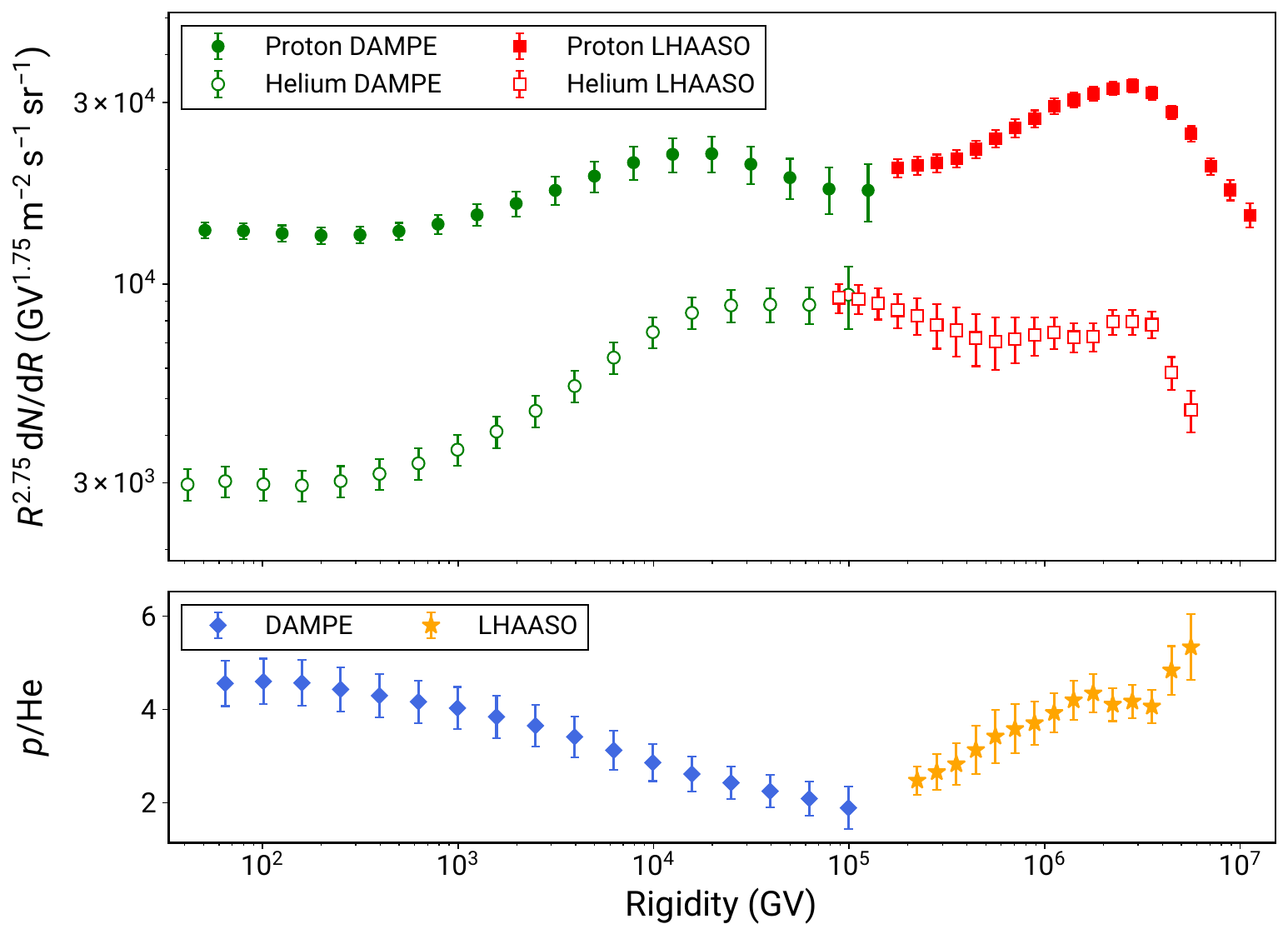}
    \caption{Rigidity spectra of protons and helium nuclei (upper panel), and the corresponding $p/{\rm He}$ ratio (lower panel). The DAMPE points are based on the latest rigidity spectra \citep{DAMPE2026}. The LHAASO points are converted from the published proton and helium energy spectra using the EPOS-LHC model \citep{LHAASOHelium2026}. Error bars show the quadratic sum of the quoted statistical and systematic uncertainties, and the uncertainties of $p/{\rm He}$ are propagated from the proton and helium flux errors. }
    \label{fig:rigidity}
\end{figure}

It is also useful to compare the DAMPE result with recent LHAASO measurements of proton and helium spectra around the knee \citep{LHAASO2025,LHAASOHelium2026}. This comparison does not simply extend the DAMPE feature to higher energy. Rather, it shows that the TeV--PeV regime contains more than one kind of structure. The DAMPE result points to a possible common sub-knee rigidity scale, while the LHAASO data show a more complex evolution of the light-component composition. These points are shown in Fig.~\ref{fig:rigidity}. The upper panel shows proton and helium
rigidity spectra. The factor $R^{2.75}$ is multiplied into the spectra only to make changes in the slope easier to see.  The filled points are protons and the open points are helium. The green points are
DAMPE direct measurements, while the red points are LHAASO air-shower measurements
converted to rigidity. The lower panel shows the proton-to-helium flux ratio, p/He. This ratio removes part of the common spectral trend and directly shows
which of the two species is harder at a given rigidity.

Before interpreting Fig.~\ref{fig:rigidity} further, it is useful to note several caveats. DAMPE and LHAASO use different detection techniques and therefore have different sources of systematic uncertainty. DAMPE is a direct-detection experiment, for which the detector response and energy unfolding depend on Monte Carlo simulations of particle interactions in the instrument. LHAASO reconstructs the primary spectra from extensive air showers, and this reconstruction depends on air-shower simulations and on the adopted hadronic interaction model. These differences should be kept in mind when the two data are compared. At the same time, the quoted systematic uncertainties of the two measurements are of comparable size in the overlapping energy ranges, and recent cross-checks indicate that the relative energy-scale difference is a few percent (as can be also seen from the error bars in Fig.~\ref{fig:rigidity}). The energy-scale offset could change the absolute flux normalization at the several-percent level, although it would not erase the spectral trends shown in Fig.~\ref{fig:rigidity}. There is also a measurement gap between the highest DAMPE proton spectrum, around 100 TV, and the lower end of the LHAASO proton spectrum around \(\sim 200\) TV. Thus the combined DAMPE and LHAASO data should be interpreted with appropriate caution, but it still provides useful information on the evolution of the light-component composition from the TeV to PeV range.

With these caveats in mind,  the upper panel of Fig.~1 shows that the proton and helium spectra both soften around 15\,TV, but do not follow exactly the same evolution beyond this rigidity. The proton spectrum shows a relatively narrow bump-like structure, followed by a higher-rigidity rise toward the PeV knee measured by LHAASO. The helium flux enhancement extends over a much wider rigidity interval. It looks more like a broad plateau than a narrow bump before the eventual softening at higher rigidity. This indicates that the helium spectral shape is not simply a rigidity-scaled copy of the proton spectrum.

In the lower panel of Fig.~\ref{fig:rigidity}, we see how the slope of the $p/{\rm He}$ flux ratio changes with rigidity, which directly reflects changes in the relative hardness of the proton and helium spectra. In the TV--PV regime, the ratio first decreases through the direct-measurement range, then turns upward as the LHAASO range is reached, flattens or turns over around the PV scale, and may rise again at the highest rigidities, 
{although this last apparent rise remains highly uncertain because of the large errors. If confirmed,} these repeated slope changes are difficult to explain with a pure propagation effect alone. A universal rigidity-dependent diffusion coefficient would affect protons and helium in nearly the same way and would largely cancel in their ratio. Species-dependent losses may modify the ratio, but their influence is not expected to be sharp at these rigidities. A more natural interpretation is that the relative contributions of source components with different chemical compositions change with rigidity. The broad helium structure in the TeV to sub-PeV range may indicate a helium-rich component, while the rising $p/{\rm He}$ ratio toward the PeV knee may point to an additional component with a larger proton fraction. This does not identify the accelerators, but it suggests that the H/He abundance ratio in the acceleration environment, or the injection efficiency of protons and helium, is relevant to the problem.

The next step is then to test the whole pattern, not only the 15\,TV break. Higher-statistics measurements of C, O, Fe, and other primary nuclei are needed to check whether the DAMPE softening is universal among primary species. Measurements of secondary nuclei are equally important, because they can separate a propagation feature from a source-related feature. If the 15\,TV softening is mainly caused by transport, related structures should appear in B/C, B/O, or other secondary-to-primary ratios. If it is mainly a source effect, the response of secondaries may be weaker or different. These spectral measurements should also be compared with CR anisotropy and gamma-ray observations of possible nearby sources.

The future High Energy cosmic-Radiation Detection facility (HERD) on board China's Space Station will be especially relevant for these tests \citep{Gargano2022HERD}. By extending direct measurements of individual CR species to PeV energies, HERD can help bridge the gap between present space-based measurements and air-shower measurements. It can test whether the apparent broad helium plateau and the turnovers of the $p/{\rm He}$ ratio between DAMPE and LHAASO are robust, smooth features, or whether additional structures appear in the poorly measured intermediate range. Its measurements of secondary nuclei such as boron can also test whether the DAMPE softening is accompanied by corresponding features in secondary-to-primary ratios expected in a propagation-based scenario.

In summary, DAMPE has reported a common softening of primary CR spectra at a rigidity of about 15\,TV. This provides important observational support for a charge-dependent interpretation rather than a simple mass-dependent one. At the same time, the combined DAMPE and LHAASO proton--helium data show that break positions alone are not enough. The helium spectrum appears to form a broader structure than the proton spectrum, and the $p/{\rm He}$ ratio changes slope several times between the TeV and PeV regimes. These composition features suggest that future models should explain not only where the spectra harden or soften, but also why the relative abundances of protons and helium change with rigidity. The key observational targets would then focus on whether the 15\,TV softening is universal, whether the same rigidity scale appears in secondary nuclei, whether the broad helium-rich component is real, and how these features connect to the higher-rigidity structures seen by LHAASO.

\bibliographystyle{unsrtnat}
\bibliography{ms}

\end{document}